%% file: main.tex
\documentclass[conference]{IEEEtran}
\usepackage{cite}
\usepackage{algorithmic}
\def\BibTeX{{\rm B\kern-.05em{\sc i\kern-.025em b}\kern-.08em
    T\kern-.1667em\lower.7ex\hbox{E}\kern-.125emX}}

\usepackage{mathptmx} 
\usepackage{amsmath,amssymb,amsfonts}
\usepackage{graphicx}
\usepackage{textcomp}
\usepackage{xcolor}
\usepackage{color,soul}
\usepackage{booktabs}
\usepackage{tabularx}
\usepackage{balance}
\usepackage{amsbsy}
\usepackage{pifont}
\usepackage{listings}
\usepackage{xspace}
\usepackage{animate}

\usepackage{fancyhdr}
\usepackage[normalem]{ulem}
\usepackage[hyphens]{url}
\usepackage[final]{microtype}
\usepackage[keeplastbox]{flushend}
\usepackage[bookmarks=true,breaklinks=true,colorlinks,citecolor=blue,linkcolor=blue,urlcolor=blue]{hyperref}

\usepackage{cleveref}
\crefformat{chapter}{\S#2#1#3}
\crefformat{section}{\S#2#1#3}
\crefformat{figure}{Figure~#2#1#3}
\crefformat{table}{Table~#2#1#3}
\crefformat{subsection}{\S#2#1#3}
\crefformat{subsubsection}{\S#2#1#3}

\fancypagestyle{firstpage}{
  \fancyhf{}
  
  \fancyhead[C]{In Proceedings of the 2024 IEEE International Symposium on Performance Analysis of Systems and Software (ISPASS)} 
  \fancyfoot[C]{\thepage}
}

\newcommand{\dci}{DCI\xspace}

\author{
Marcelo Orenes-Vera, Esin Tureci, Margaret Martonosi, David Wentzlaff\\
Princeton University \\
\{movera, esin.tureci, mrm, wentzlaf\} @princeton.edu
}
\newcommand{\repository}{repository\cite{muchi_repo}\xspace}
\newcommand{\repo}{repo\cite{muchi_repo}\xspace}
\newcommand{\proj}{MuchiSim\xspace}
\title{\proj: A Simulation Framework for Design Exploration of Multi-Chip Manycore Systems}

\begin{document}
\maketitle
\thispagestyle{firstpage}
\pagestyle{plain}

\newcommand{\subf}[2]{%
  {\small\begin{tabular}[t]{@{}c@{}}{\tiny}
  #1\\#2
  \end{tabular}}%
}

\input{tex/muchisim.tex}

\section*{Acknowledgements}
This material is based upon work supported in part by National Science Foundation (NSF) award No. 1763838, and based on research sponsored by the Air Force Research Laboratory (AFRL) and Defense Advanced Research Projects Agency (DARPA) under agreement FA8650-18-2-7862.
In addition, Martonosi received separate NSF support while serving at NSF as an IPA rotator.~\footnote{The U.S. Government is authorized to reproduce and distribute reprints for Governmental purposes notwithstanding any copyright notation thereon. The views and conclusions contained herein are those of the authors and should not be interpreted as necessarily representing the official policies or endorsements, either expressed or implied, of NSF, AFRL, DARPA or the U.S. Government.}


\bibliographystyle{IEEEtranS}
\bibliography{refs}

\end{document}

%% file: tex/muchisim.tex
\begin{abstract}

The design space exploration of scaled-out manycores for communication-intensive applications (e.g., graph analytics and sparse linear algebra) is hampered due to either lack of scalability or accuracy of existing frameworks at simulating data-dependent execution patterns.
This paper presents MuchiSim, a novel parallel simulator designed to address these challenges when exploring the design space of distributed multi-chiplet manycore architectures.
We evaluate MuchiSim at simulating systems with up to a million interconnected processing units (PUs) while modeling data movement and communication cycle by cycle.
In addition to performance, MuchiSim reports the energy, area, and cost of the simulated system. It also comes with a benchmark application suite and two data visualization tools.

MuchiSim supports various parallelization strategies and communication primitives such as task-based parallelization and message passing, making it highly relevant for architectures with software-managed coherence and distributed memory.
Via a case study, we show that MuchiSim helps users explore the balance between memory and computation units and the constraints related to chiplet integration and inter-chip communication.
MuchiSim enables evaluating new techniques or design parameters for systems at scales that are more realistic for modern parallel systems, opening the gate for further research in this area.
\end{abstract}

\begin{IEEEkeywords}
graph, sparse, communication-intensive kernels, simulator, design-exploration, multi-chip, scale-out, manycore
\end{IEEEkeywords}

\vspace{+1mm}
\section{Introduction}
\vspace{+1mm}

The end of Moore's law and the increase in application dataset sizes have driven the recent surge in development of scale-out manycore systems~\cite{cerebras,groq,tesla_dojo,tpu_google,nvidia_a100,manticore,graphcore,tenstorrent}.
Design-space exploration for these architectures at large scales employs roofline and analytical models to estimate performance, by relying on the deterministic nature of communications as well as computations for many ML and dense applications \cite{rooflineModel}.
However, communication pattern in data-dependent communication-intensive (\dci) applications, such as graph analytics, sparse linear algebra, and database operations, is not deterministic, and therefore estimation of performance for running these applications on a system under design requires simulation tools.
Current simulators cannot evaluate \dci applications at scale due to the limited scalability of detailed processing unit models.
Thus, recent work on architecture designs for \dci applications are evaluated on relatively small scales, and their network design---whose performance impact would be exacerbated at larger scales---is not well explored~\cite{polygraph,maple,fifer,pipette,graphicionado,flexagon,taskstream,graphattack}.

Current manycore architectures typically emphasize accelerating computation, dedicating significant chip area to it, at the expense of memory and network resources. 
However, when the workload is highly parallelized, data movement increasingly becomes the limiting factor for performance.
This is especially important when using such manycore systems to accelerate communication-intensive applications such as graph analytics, sparse linear algebra, and spectral methods~\cite{dalorex,cerebras_fft}, which necessitate a careful balance between computational, memory and network resources.
Thus, in manycore systems with thousands to millions of cores, factors such as the number of cores per silicon die, the number of dies per package, memory hierarchy, network-on-chip (NoC) and inter-chip interconnect, become critical design considerations.
The inability to evaluate these aspects at a large scale creates significant gaps between designed systems and their actual realizations.

To address these issues, we introduce \proj\footnote{\proj is derived from the Spanish word \textit{muchisimo} (very much), alluding to its ability to simulate a large number of PUs in a multi-chip setting. \proj is pronounced as "moo-chee-sim".},
a parallel simulator for exploring the manycore architecture design space, with a special focus on communication-intensive applications\footnote{At large scale, memory-intensive applications lead to a communication bottleneck; we call this category of applications communication-intensive.}.
\proj scales up to millions of interconnected PUs by relying on user instrumentation for code runtime on the compute units while simulating data movement and communication every cycle.
It also includes energy, area and cost models for multi-chip distributed systems; these are critical to reason about design tradeoffs.
In addition, \proj includes an application benchmark suite and data visualization tools to facilitate the analysis of its outputs.

As industry trends towards explicit data movement for large manycores due to the prohibitive cost of hardware-based coherence~\cite{cerebras,groq,tesla_dojo, manticore}, \proj becomes particularly relevant.
\proj is applicable for architectures with distributed memory and software-managed coherence, which is becoming increasingly popular in modern scale-out systems.
\proj facilitates exploring the integration granularity and balance between memory and computation units, as well as constraints related to chiplet integration and inter-chip communication.

With a multifaceted analysis aimed at expanding the current understanding of manycore architectures for \dci applications, \proj offers a flexible yet powerful framework for exploring many design configurations, thereby contributing to the design and optimization of future large-scale systems.
\vspace{+1mm}

\textbf{Our main technical contributions are}:
\begin{itemize}
\itemsep0em 
\item A novel performance modeling approach for distributed manycore architectures that allows for scalable simulations of data-dependent and communication-intensive applications across millions of PUs.
\item Performance, energy, area and cost modeling of multi-chip module (MCM) and interposer-based integrations, critical in the design of scalable manycore systems.
\item Support for different parallelization strategies (do-all and task-based) and communication primitives (e.g., message-passing and reduction trees).
\item A benchmark suite of eight applications especially programmed for distributed scale-out systems.
\item Visualization tools that allow comparing system-wide metrics for different evaluations (i.e., design configurations, applications and datasets) and analyzing per-PU metrics throughout the entire execution of a particular evaluation.
\end{itemize}

\textbf{We evaluate \proj and demonstrate that:}
\begin{itemize}
\itemsep0em
\item \proj is the first open-source framework that precisely simulates \dci applications with billion-element datasets parallelized across a million PUs within tens of hours.
\item Its parallelization achieves linear speedups up to a host thread count equal to the number of columns of the manycore grid being simulated.
\item \proj closely matches runtime and area of the real runs of the Cerebras Wafer-Scale Engine when using their reported workload implementation and network specification.
\item \proj is a powerful design-exploration tool that helps identify optimal configurations for different metrics (e.g., performance-per-dollar) and applications.
\end{itemize}

The rest of the paper is organized as follows:
\cref{sec:background} provides background on existing simulation approaches and motivates our work;
\cref{sec:approach} details the architecture class that \proj simulates, the design of \proj and capabilities, and the benchmark and visualization tool that it includes;
\cref{sec:results} presents the validation and scalability analysis of \proj, in addition to a case study that uses \proj to study different memory integrations;
and \cref{sec:conclusions} concludes the paper and discusses future work.
\vspace{-2mm}
\section{Background and Motivation}\label{sec:background}

As the number of processing units (PUs) used to parallelize a workload increases, the network bandwidth and topology become critical factors in the performance of the system.
Depending on the architecture, bottlenecks may also shift or worsen with the level of parallelism.
Thus, detailed simulation of large manycore systems is crucial for design-space exploration, to avoid making suboptimal design decisions that may be hard to correct later in the design process.

The computer architecture community has seen a lot of promising research on hardware-software co-designs and optimizations for data-dependent communication-intensive (\dci) applications~\cite{prodigy,maple,graphicionado,graphpulse,ozdal,chronos, pipette, fifer, polygraph, jeffrey_hive, jeffrey_swarm,cohort,dalorex,dcra,spatula}.
However, these ideas are often evaluated on a limited number of PUs (hundreds to a few thousand) or relatively small datasets (up to millions of data elements), as increasing either of these dimensions significantly increases simulation time.
\proj enables the exploration of architectural ideas on larger scales, as well as offers key parameters to explore the balance of hardware resources and power allocated to memory, network and compute.

Maximizing accuracy as well as performance is the main objective for the design and engineering an architecture simulator, as it dictates reliability as well practicality of design experiments~\cite{Zarrin2017}.
For the rest of this section, we outline the critical features of a scale-out manycore system simulator (and their potential trade-offs) and review existing simulators.

\subsection{Full-system vs. Application-level Simulation}

Full-system simulators offer the ability to evaluate the performance of a target design in the context of other system components as well as a complete software stack including the operating system (OS).
However, this comes at a great expense of speed and scalability even for small-scale systems.
Nevertheless, when the execution significantly relies on OS and I/O processes, full-system simulators are essential when evaluating the performance of a design.
Simulators such as GEM5, SimFlex, COTSon, RAMP Gold are some of the most commonly used full-system simulators with varying scope and capabilities~\cite{gem5,simflex, cotson, ramp-gold}.

In the manycore architectures we are considering in this study, PUs are not expected to run the OS but rather behave as accelerator processing elements.
These manycores save silicon area and power by not implementing hardware coherence and instead having an explicit view of the memory.
In addition, PUs rely on software to orchestrate the data movement---optimized based on application specifics~\cite{cerebras, dalorex, tascade, sambanova,tenstorrent}.
This makes an application/user-level simulator for these systems more appropriate.

Application-level simulators can scale to higher degrees of parallelism than full-system simulators by simulating up to a couple of thousand PUs~\cite{graphite, sanchez2013zsim, prime, mcsimA, cotson}.
However, existing simulators cannot efficiently simulate systems beyond this size, and some of them lack detailed network modeling, which significantly affects the accuracy of larger system simulations.
BigSim~\cite{bigsim} is a simulator originally developed to simulate Blue Gene that performs detailed network simulation after the emulation of the program on a real system to account for network traffic, achieving scaling in tens to hundreds of thousands of cores.
BigSim relies on the program to be deterministic, but certain \dci workloads like graph algorithms do not converge deterministically unless additional synchronization steps are employed.

\subsection{Simulating \dci Applications on Manycores}

This paper focuses on a class of tiled, distributed manycore architectures.
This class of architectures is composed of an interconnected number of machine nodes, each with a board of chip packages, further subdivided into chiplets and processing tiles.
Therefore the entire compute system can be viewed as a hierarchically connected grid of tiles.

\proj is a framework to explore the design space of this class of architectures, with a focus on \dci applications.
The challenge with these applications is that their data-dependent execution requires functional simulation to precisely capture performance and energy usage.
When parallelizing \dci applications across thousands of PUs and billions of data elements, faithfully simulating all architecture components every cycle becomes infeasible.

\textit{\textbf{PU modeling:}}
Since prior literature has identified the memory bandwidth and inter-PU communication to be the main bottlenecks of these applications, \proj focuses on modeling the network and memory systems, while executing compute tasks natively on the host and relying on the user to provide a performance model for the compute time (e.g., cycles per basic block).
A similar approach has been previously used by PriME~\cite{prime}, but PriME simplifies the PU model by simulating each instruction as 1 cycle, whereas \proj can use that or a detailed user-provided model (\cref{sec:approach}).
Because the PUs used in scale-out manycores do not interact with the OS and have a simple view of the memory system, it is possible to model the execution using user-provided cycle counts and without detailed core simulations.
Simulators such as SimpleScalar~\cite{simpleScalar}, SimFlex~\cite{simflex}, and Gem5~\cite{gem5} offer detailed core modeling with a wide range of supported ISAs.
In addition, simulators such as MosaicSim~\cite{mosaicSim} and SimTRaX~\cite{SimTRaX} provide LLVM-based instruction sets.
When precise PU models are desired, these simulators can be used to provide the task-specific cycle counts to be incorporated into \proj simulations.

\textit{\textbf{Network modeling:}}
Since inter-PU communication becomes a significant performance factor for large-scale manycore systems, the capability to model different configurations of the network-on-chip (NoC) and the inter-chip interconnect in detail is essential when modeling these systems.
While there are several cycle-accurate NoC simulators such as Xpipes~\cite{xpipes}, NOXIM~\cite{noxim}, SICOSYS~\cite{SICOSYS}, BookSim~\cite{booksim}, among others~\cite{horro}, most of the existing multicore system simulators~\cite{dec++,prime,graphite} do not model the NoC in detail.
Although some implementations incorporate these NoC models within system simulators (e.g. Xpipes with Gem5), a functional architecture simulator (with cycle-accurate NoC modeling) able to simulate beyond a thousand PUs has not been offered so far.
\proj fills this gap.

\subsection{Scope of Applicability}

The class of architectures for which \proj is applicable includes---but not limited to---tiled accelerators~\cite{dalorex, tascade_massive, graphq, tesseract, polygraph} and dataflow machines~\cite{cerebras, sambanova, groq, tenstorrent,codelet}.
\proj can also be used to simulate systolic array architectures like Google's TPU~\cite{tpu_google}, but these designs usually execute deterministic workloads with near-neighbor communication, and thus they would be modeled much faster using analytical models.

Unlike general-purpose simulators for HPC such as SST~\cite{sst}, focusing on this architecture class allows \proj to be a highly-scalable, parallel, light-weight simulator while providing detailed simulations of critical aspects of \dci workloads such as NoC modeling at flit-level granularity.
\proj could also potentially be used to model the performance of workloads running on more general-purpose manycores~\cite{manticore, cifer, decades, celerity} and server-class chip~\cite{esperanto, ampere, amd_epyc_isca}, provided that the application is written in a way that does not require hardware coherence.

\section{\proj Simulation Framework}\label{sec:approach}

In \cref{sec:ap_hw_components} we first present the hardware components of \proj's target architecture class, and how they are hierarchically organized.
Then, \cref{sec:ap_application} explains the execution models supported in \proj and how to describe applications.
Next, \cref{sec:ap_perf} elaborates on how \proj simulates the execution of the applications on the target architecture.

\proj is a parallel simulator written in C/C++ which does not use external libraries beyond threading, and thus, it can be compiled with any C++ compiler that includes OpenMP (\texttt{-fopenmp}) or pthreads support (\texttt{-lpthread}).
The code can be found in our \repository under the \texttt{src} folder, where the \texttt{configs} subfolder contains the files to configure the system to simulate, and its latency, energy, area and cost parameters (described in \cref{sec:ap_energy_model} and \cref{sec:ap_cost}).
The \texttt{gui} and \texttt{plot} folders contain the visualization tools to analyze and compare simulation logs (\cref{sec:ap_visualization}).
The \texttt{apps} and \texttt{datasets} folders contain the benchmark suite of applications and datasets (\cref{sec:apps}).
In addition, the repository includes a Readme file explaining how to configure and run \proj for different experiments.

\subsection{The Target Architecture Class}\label{sec:ap_hw_components}

\begin{figure*}[t]
\centering  
\includegraphics[width=0.98\textwidth]{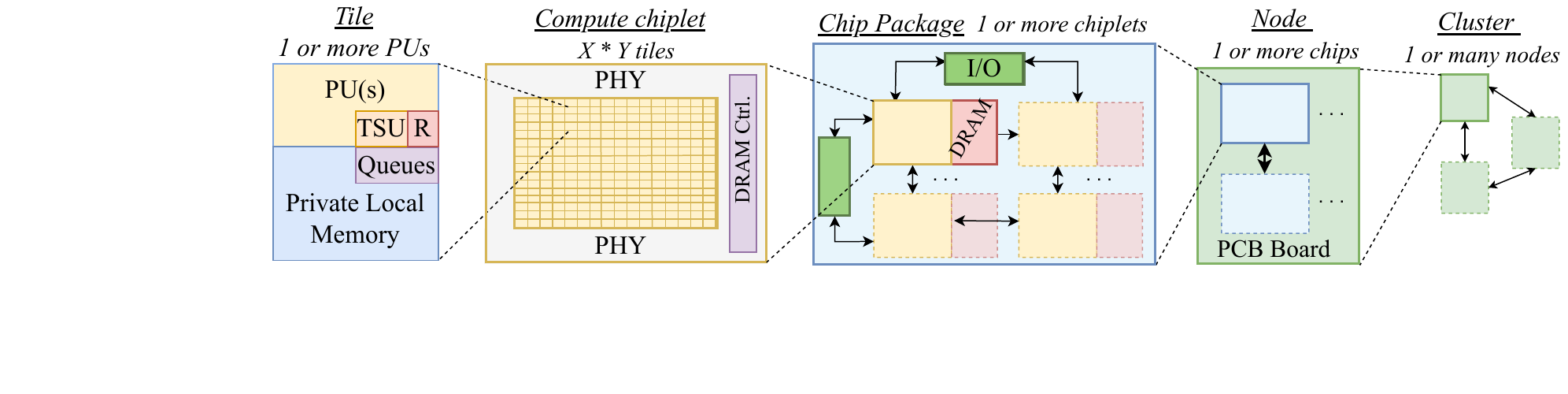}
\vspace{-3mm}
\caption{
Hierarchical overview of how tiles can be organized on \proj's target architecture.
The board of a cluster node may contain one or multiple chip packages, each with one or multiple chiplets.
Packages can be composed of only compute chiplets (with a grid of tiles), or also DRAM chiplets (adjacent to the compute chiplets).
Chiplets also include the physical layer (PHY) for inter-chiplet communication.
A tile contains one or more processing units (PUs), a private local memory (PLM), a network router (R), and a task scheduling unit (TSU).
The task queues are mapped into the PLM.
} 
\vspace{-1mm}
\label{fig:hierarchy}
\end{figure*}

Throughout the paper, we refer to the simulated architecture as the \textit{design under test} (DUT).
We call the machine that runs the simulator, the \textit{host}. We use \textit{DUT's host} to refer to the machine for which the DUT may behave as an accelerator.

\cref{fig:hierarchy} shows the hierarchy of \proj's target architecture class.
A cluster is composed of interconnected boards ($\sim$nodes) that contain one or more chip packages ($\sim$sockets).
Each package is composed of one or several compute chiplets of any size (from a few mm$^2$ to as big as a wafer), optionally integrating DRAM chiplets with the compute chiplets.
Each compute chiplet has a grid of tiles, where each tile contains one or more PUs, a private local memory in SRAM, a router, and a task scheduling unit (TSU).

\textit{\textbf{Processing Units (PUs):}}
A PU can be an ISA-programmable core, a CGRA unit, or a hardware accelerator, depending on the performance, energy and area models specified.
We have only evaluated cases where the PUs are homogeneous across the chiplet, but \proj could potentially simulate heterogeneous manycores~\cite{decades,sambanova,cifer_cicc} since a different performance model could be provided for different PU IDs.

\textit{\textbf{Network-on-chip (NoC):}}
\proj supports evaluating 2D mesh and folded torus topologies\cite{topology_study} (with dimension-ordered routing) for the NoC that connects the routers of every tile. This NoC can span multiple chiplets and chip packages, up to the level of a cluster node.
Several physical NoCs can be evaluated, with the same or different NoC widths.
Additional NoCs can be regular (i.e., connecting all routers) or Ruche~\cite{ruche_batten,ruche_taylor} (i.e., connecting to one for every $R$ routers).

\textit{\textbf{Routers}} generally have five bidirectional ports: North, South, East, West, and the PU ports (connected to the PU's input and output queues).
When configuring hierarchical inter-chiplet~\cite{dcra} or Ruche NoCs, some or all routers have an additional set of cardinal ports (for a total of nine).
\proj can also evaluate the Tascade router support for asynchronous and opportunistic reduction trees~\cite{tascade}.

\textit{\textbf{Task Scheduling Unit (TSU):}}
To support task-based parallelizations and message-triggered tasks (see \cref{sec:ap_application}), \proj considers a TSU per tile, which takes messages from the input queues (one per task type) and schedules them to a PU.
The TSU decides which task to schedule next---from the ones available at the queues---based on a scheduling policy.
The policies currently supported are round-robin, priority-based (i.e., prioritizing tasks from a particular queue), or occupancy-based (to prevent full queues from backpressuring the network).

\textit{\textbf{Private Local Memory (PLM):}}
The size of each tile's memory is specified in Kibibytes (KiB); it is the same for all the tiles.
\proj supports using the PLM as a write-back cache, when DRAM is integrated on-package, or as a scratchpad (given the tile's local address space) when the chip's main memory is the tile-distributed SRAM.
In cache mode, we model the area and energy of the tags (and valid/dirty bits) using part of the local SRAM; the width of a cacheline is equal to that of the bitline of the DRAM memory controller (set to 512 by default).
Cache misses are directed to the on-chip memory controller, which fetches the full cacheline from DRAM directly since there is no hardware coherence in our target DUT.
Dirty lines are written back to DRAM upon eviction.

\textit{\textbf{Queues:}} Inputs and outputs for message triggered tasks (see \ref{sec:ap_application} for details) are mapped into the PLM of each tile and currently modeled as circular FIFOs, which allows for compile-time configuration on the DUT.
Thus, in our energy model, queue reads and writes are similar to load and store operations.
Alternatively, one could change the model to consider them as separate structures, with the corresponding tradeoff in area and energy.

\textit{\textbf{Prefetching:}}
When DRAM is present in the design, \proj supports modeling next-line and pointer-indirection data prefetching from DRAM into the PLM (used as a cache).
The latter is enabled by the fact that tasks can be split at pointer indirection and the TSU can prefetch the data that tasks (waiting in the input queue) will use.

\textit{\textbf{Chiplet Integration}}:
The network routers at the edges of the compute chiplets are connected to the PHYs for inter-chiplet communication~\cite{die2die_comp}.
When DRAM is integrated on-package, \proj models the pairs of DRAM and compute chiplets to be connected via a silicon interposer, while these interposers lay on an organic substrate (details in \cref{sec:ap_energy_model}).
Alternatively, \proj can model the compute chiplets to be connected via a single silicon or organic interposer.
In our model, the choice of the interposer mostly affects maximum communication bandwidth, area of the PHY, and energy per bit (see \cref{table:wire_param}).

\textit{\textbf{Interconnect links:}}
\proj offers parameters to set the width and the number of interconnect links across chiplets (within a package) and across chip packages.
Across cluster nodes, one can also set the multiplexing factor of the links between nodes, i.e., how many edge tiles share an inter-node link.
To interconnect nodes, \proj currently supports evaluating a mesh topology, but this could be extended to support other cluster-level topologies~\cite{slimf,fattree,dragonfly,benner2012optical,blue_gene}.

\subsection{Describing and Mapping an Application}\label{sec:ap_application}

The code for an application and its potential DUT compile-time configurations are described as C/C++ header files in a structured manner using a template that defines a series of functions.
The file for the application to simulate (\texttt{src/apps}) is included via preprocessor macros so that the simulator code is compiled together with the application code.

\textit{\textbf{Configuration functions:}}
The functions with the prefix \texttt{config\_} are invoked only once at the beginning of the simulation.
They allow overriding software parameters of the DUT for which the configuration files have set default values.
These parameters include the sizes of the queues per task, as well as the parameters for prioritizing the scheduling of certain task IDs.
Other software parameters of the DUT are macros that need to be set when compiling the simulator.
\proj uses pre-processor macros in some places to save branches and improve performance.

\textit{\textbf{Address space and dataset layout:}}
\proj maps the memory address space contiguously such that the PLM of each tile is assigned a chunk of this contiguous space. 
The dataset arrays for a given application are allocated and mapped to memory across tiles (at \texttt{config\_app}) depending on how a user or mapping technique~\cite{affinalloc} decides so.
On our application suite, the dataset is scattered so that each tile has an equal chunk of each data array.

\textit{\textbf{Message-triggered tasks (MTTs):}}
The functions associated with an identifier (e.g., task1) describe the different types of MTTs that are invoked upon receiving a message into their corresponding input queue.
Each MTT is associated with an IQ matching its identifier (ID).
Defining MTTs enables simulating communication primitives like message passing, active messages, and non-blocking remote procedure calls (RPCs).
Tasks can invoke other tasks by placing messages directly into another task's input queue (IQ)---if the task is going to be executed on the same PU---or into a channel queue (CQ)---if the task is meant to be executed on a different PU.
Each CQ drains into its logical channel in the network, and each channel is associated with an IQ ID.
To avoid network deadlocks, loops in the association between MTTs are not allowed (e.g., task1 invoking task2 and task2 invoking task1).
In other words, the dependency chain must end on a leaf task (a task not invoking other tasks); a new phase of computation can start after a local or global barrier.

\textit{\textbf{Init task:}}
The function with the suffix \texttt{\_init} is invoked once at the beginning of each kernel.
The \texttt{kernel\_count} variable can be used to distinguish behavior between different kernels.
Multiple kernels can be simulated in sequence (with global barrier synchronization between them) to compose an application.
The combination of the Init task and MTTs per kernel enables the simulation of a variety of programming models. However, recursion is not supported.

\textit{\textbf{Supported parallelization modes:}}
On the one extreme, the Init task can be empty if the entire kernel only comprises MTTs invoking each other, starting from a seed invocation coming from the DUT's host.
On the other extreme, a kernel can be entirely included in the Init task (not invoking remote tasks).
All task functions can access the identifier of the PU for which the task is being simulated and the size of the grid of PUs, and thus, kernels can be parallelized across PUs similarly to using \emph{pthreads}.
\cref{sec:ap_perf} describes how \proj iterates over every PU to check whether there are tasks ready to execute.

\textit{\textbf{Result-check function:}}
Since \proj is a functional simulator, the application results can be compared with reference values.
The \texttt{compare\_out} function can be used to compare the outputs directly or write them into a file to be compared outside the simulator.

\subsection{Simulating the Application Runtime}\label{sec:ap_perf}

Although the tiles of the DUT may be organized hierarchically, for parallelization purposes, the simulator considers a global grid of tiles, of which each host thread simulates a slice of one or more columns.
There are two types of threads, the ones that simulate the TSU and the PU tasks (\emph{execution threads}) and the ones that simulate the network (\emph{router threads}).

The \textit{\textbf{execution threads}} iterate over every tile in their slice of the grid, checking if there are tasks ready to be executed in the tile's IQs, in addition to the Init task.
If there are, the TSU selects the next task type to execute based on the configured policy.
Then, the function associated with that task is executed, and the task delay is obtained from the latency-instrumented code.
The codes included in our application suite are instrumented considering the execution of simple in-order PUs.
However, users could change that to reflect other PU models.
For memory operations, \proj offers a special \texttt{dcache} function that returns the latency to fetch a given memory address, depending on whether it hits in the data cache, and the configured memory system.
The function call simulating the task may have pushed messages into other IQs or CQs, with the corresponding timestamps, so they cannot be routed until the router thread is executed for that timestamp.
Once the task function returns, the clock for that PU advances as many cycles as the task delay, and the execution thread continues evaluating the next PU.
The execution thread keeps iterating over the PUs until there are no more tasks to be done, and the network is empty (i.e., no more messages to be routed).
This models a hardware-based termination condition based on idleness, and by default, we set its latency to two times the diameter of the network.

The \textit{\textbf{router threads}} iterate every cycle over every router in their slide of the grid.
For every input port, they check whether there is a message to be routed and a free buffer slot in the destination port (the buffer size is also parameterizable).
When two or more input ports want to route to the same output, the priority is determined via a round-robin policy (static priority could also be configured).
A message is only routable if the timestamp (set at injection and updated every hop) is less than or equal to the current router cycle; the timestamps do not exist in the DUT, but they are used to allow PUs and routers to be simulated in parallel.
The synchronization between the execution and router threads is done based on their wall clock time, i.e., a router can never be simulated ahead of its tile's PUs.
In addition, a PU cannot start executing a new task until the router thread has caught up with the clock time of the slowest PU in the tile.
Thanks to this type of synchronization, \proj can seamlessly support different frequencies for the PUs and the network, with any ratio between them.

\textit{\textbf{Frequency:}}
The DUT configuration file in \proj distinguishes between the peak frequency supported by the design and the operating frequency (at which the DUT is evaluated) for the PUs and the NoCs.
Peak design frequency is going to affect the silicon area of the design, while the operating frequency affects the power consumption due to voltage scaling. The default model for this is described in \cref{sec:ap_energy_model}).
Peak and operating frequencies can be changed independently for PUs and NoCs, which is useful to evaluate architectures looking to support applications with different compute and network demands. 
For example, having a higher peak PU frequency allows raising its operating frequency when needed (e.g., to increase compute capacity) while lowering it to save power for applications with low arithmetic intensity.
The configuration file has by default a peak and operating frequency of 1 GHz for all components.

\textit{\textbf{SRAM model:}}
The default latency, area and energy parameters of the SRAM memories are based on 7nm technology at 1 Ghz~\cite{renesas_ff7nm} (see \cref{table:wire_param}).
Unlike the PU and NoC, SRAM memories have by design a narrower operating voltage and frequency range, therefore we do not support changing these and we only use the default values for which we have a reference (\cref{table:wire_param}).
\proj models scaling the size of SRAM memories by increasing the number of banks.
Only the active banks consume static energy, and we model the energy of the multiplexor tree that selects the bank to grow by 50\% at each doubling step.
The access latency is also modeled to increase by 1 nanosecond at quadrupling steps beyond 512 KiB.
All these modeling options and values can be changed in the simulator parameters.
On the DUT configurations with multiple PUs per tile, the SRAM is shared among them, but bank conflicts are not modeled yet for this case.

\textit{\textbf{DRAM model:}}
Each compute chiplet may be paired with a memory device of a given number of channels.
By default, \proj considers an HBM2E device with eight 64GB/s channels.
On the 2.5D integration shown in \cref{fig:hierarchy}, the memory controller sits on one edge of the chiplet, adjacent to the closely integrated DRAM device.
The bus connecting the tiles to the memory controller is separate from the task-communication network.
Since each memory channel is shared by many tiles, the contention is modeled by imposing that the memory channel can only take one request per cycle, and keeping the count of the transactions of each channel.
For example, if a request is done at cycle $X$, but the memory channel has received $Y$ transactions (where $Y > X$), then the delay if this request is $Y - X + $ the round-trip to the memory channel. 
Increasing the capacity of HBM increases the cost linearly (see \cref{sec:ap_cost}) but not the device area since it is a 3D stack. 
The HBM bandwidth available to the processing units can be defined by changing the number of channels or devices integrated with each compute chiplet, or by changing the number of tiles or PUs per tile of these chiplets.

\subsection{Energy and Area Model}\label{sec:ap_energy_model}

\cref{table:wire_param} summarizes some of the energy, latency, and area parameters for communication links and memory technology, while the full set of default energy, area and performance parameters of \proj can be found in our \repo.
The simulator periodically logs performance and energy during the simulation, in what we call \textit{frames}, at a rate that can be configured.
This name will become more apparent in \cref{sec:ap_visualization}, which describes the visualization tools.
Recording frames allows the user to observe the progress of the simulation, by visualizing several metrics (per PU or averaged) for each frame and aggregating until that frame.
At the end of the simulation, the simulator also outputs statistics over metrics like throughput, average power, and network traffic at different levels of the hierarchy.

The files starting with \texttt{calc\_} under the \texttt{src/common} folder contain the functions that calculate the performance metrics, energy, area and cost of the different DUT components based on the simulation of the application execution.

Because there is often no ground truth for energy and cost parameters but rather estimations or assumptions,
\textbf{\textit{\proj allows post-processing a given simulation to re-calculate the energy and cost with different model parameters}}.
In addition to the execution log file, \proj creates a separate file with many performance counters collected during the simulation process, such as messages hops and memory accesses at different levels, and instructions executed for each type.
The counters file is then provided as an argument to our separate post-processing executable---which uses the same configuration and parameter files as the simulator, with potentially new values---to re-calculate the energy usage, and the DUT area and cost.

The model for how area and voltage grows with peak and operating frequency can be re-evaluated during post-processing.
By default, \proj simplistically models the area of the PUs and routers to grow by 50\% of the increase in their peak frequency.
This model can be refined by synthesizing particular RTL components with different peak frequencies and measuring their area.
A simulation can be post-processed with a new area model to re-calculate energy and cost.

For our \textbf{\textit{voltage scaling model}}, we fit a ridge regression to the frequency and voltage data from the shmoo plots of chips with 5, 7 and 12 nm transistor nodes~\cite{5nm_shmoo,7nm_shmoo,decades}.
The current model grows voltage with $0.06 + 0.13 * freq + 0.06 * node$, which could be adjusted by adding data to \texttt{src/voltage\_model.py}.

\textit{\textbf{DRAM integration:}}
The standard integration of DRAM and compute chiplets in \proj is a 2.5D integration via a passive interposer~\cite{nvidia_a100,shapphire_rapids}.
However, since the adjacent compute chiplet has dedicated access to the DRAM device, another possible integration would be having the DRAM device on top of the compute chiplet, which would need to be redimensioned to behave as an active interposer, in addition to posing a power and thermal challenge~\cite{amd_die_stacking,die_stacking}.
\proj supports studying this mode by adjusting the area, cost, and wire energy with the DRAM integration, but we do not consider a different latency.
\proj reports power density, which can be used to estimate the thermal feasibility of a 3D integration.

\textit{\textbf{Chiplet PHY and Package I/O}} are affected by the width, frequency and topology of the on-chip and off-chip network, respectively.
\proj configures one physical network by default, but up to three independent NoCs have been evaluated (one for each of the task types of the benchmark suite, \cref{sec:apps}).

\begin{table}[t]
\centering
\small
\caption{Default values of energy (E), bandwidth, latency, and area parameters of links and memory devices modeled in \proj.}
\begin{tabularx}{\columnwidth}{@{\hspace{0pt}}l@{\hspace{-30pt}}r@{\hspace{0pt}}}
\toprule
\textbf{Memory Model Parameters} & \textbf{Values} \\
\midrule
SRAM Density\ & 3.5 MB/mm$^2$~\cite{renesas_ff7nm} \\
SRAM R/W Latency \& E. & 0.82 ns \& 0.18 / 0.28 pJ/bit~\cite{renesas_ff7nm} \\
Cache Tag Read \& cmp. E. & 6.3 pJ~\cite{renesas_ff7nm,ariane_cost}\\
HBM2E 4-high Density\ & 8GB on 110mm$^2$ (75 MB/mm$^2$)~\cite{hbm2_sk_hynix} \\
Mem.Channels \& Bandwidth & 8 x 64GB/s~\cite{hbm2_sk_hynix} \\
Mem.Ctrl-to-HBM Latency \& E. & 50 ns \& 3.7 pJ/bit~\cite{fine_grain_dram,hbm_samsung_power} \\
Bitline Refresh Period \& E. & 32 ms \& 0.22 pJ/bit~\cite{refresh_time_hbm,dram_activation_energy} \\
\midrule
\textbf{Wire \& Link Model Parameters}  & \textbf{Values} \\
\midrule
MCM PHY Areal Density                   & 690 Gbits/mm$^2$~\cite{die2die_comp} \\
MCM PHY Beachfront Density              & 880 Gbits/mm~\cite{die2die_comp} \\
Si. Interposer PHY Areal Density        & 1070 Gbits/mm$^2$~\cite{die2die_comp} \\
Si. Interposer PHY Beachfront Density   & 1780 Gbits/mm~\cite{die2die_comp} \\
Die-to-Die Link Latency \& E.          & 4 ns \& 0.55 pJ/bit ($<$25 mm)~\cite{bow} \\
NoC Wire Latency \& E.              & 50 ps/mm \& 0.15 pJ/bit/mm~\cite{pim_hbm} \\
NoC Router Latency \& E.            & 500 ps \& 0.1 pJ/bit \\
I/O Die RX-TX Latency                   & 20 ns~\cite{pcie6} \\
Off-Package Link E.                 & 1.17 pJ/bit (upto 80mm)~\cite{nvidia_chiplets}\\
\bottomrule
\end{tabularx}
\label{table:wire_param}
\end{table}

\subsection{Cost Model}\label{sec:ap_cost}

We also added a fabrication cost model to the simulator to study the cost-effectiveness of different architecture configurations.
Similarly to the energy model, the cost model is decoupled from the runtime simulation process, i.e., cost and energy can be re-calculated post-simulation for different parameters.
This is useful to study how price variations can change the cost-effectiveness of different DUT configurations.

\textbf{\textit{Silicon cost:}}
By default, we consider 7 nm technology for transistors; we assume that a 300 mm wafer with this transistor process costs \$6,047~\cite{lithovision}.
We obtain the cost per die by dividing the wafer cost by the number of good dies, which we calculate using 0.2 mm scribes, 4 mm edge loss, and 0.07 defects per $mm^2$.
We integrate and validate~\cite{yield_calc} die yield calculations in our cost model using Murphy's model.
When comparing the cost-effectiveness of the simulated architecture, we do not include the non-recurring engineering cost of the compute dies since all the options use the same technology.

\textbf{\textit{Packaging cost:}}
We model the cost of the 65 nm silicon interposer connecting a compute die with a DRAM device (including bonding) to be 20\% of the price of a compute die~\cite{cost_model_package}, and the cost of an organic substrate to be 10\% of the price of an equal-sized compute die, and the bonding to add an extra 5\% overhead~\cite{cost_model_interposer,bonding_yield}.

\textbf{\textit{DRAM cost:}}
While the cost of HBM2 is not disclosed, we made an educated guess using public sources~\cite{lithovision,cost_hbm}.
By default, the cost model assumes \$7.5/GB, which is more affordable than when HBM was first released in 2017.
One could expect this price to decrease over time as more vendors fabricate HBM or the process matures~\cite{micron_hbm,hbm2_sk_hynix,hbm_samsung,hbm3_sk_hynix}.
\proj's post-processing tool allows evaluating the performance-per-dollar of a given simulation in the light of different DRAM cost scenarios.

\subsection{Visualization Tools}\label{sec:ap_visualization}

\proj comes with two data visualization tools: a command-line interface (CLI) tool to generate plots including multiple application executions, and a graphical user interface (GUI) tool---coded in PyQt5---to visualize \proj metrics over time for a single execution.
Both tools are written in Python and use matplotlib to generate the plots.

The CLI tool allows \textbf{plotting multiple metrics for combinations of DUT configurations and sizes, and different applications and datasets}.
These metrics include runtime, throughput (FLOPS or TEPS), energy (and its breakdown), cost (in USD), simulator time, arithmetic intensity (FLOPS divided by data loads or network traffic), overall network traffic (in message hops), and cache hit-rate, among others.
They can be plotted as absolute numbers (as in \cref{fig:sim_time}) or normalized to a baseline configuration (as in \cref{fig:use_case_mem}).

The GUI tool allows \textbf{visualizing performance counters throughout the application execution}, such as router port collisions and utilization, end-point contention, PU utilization, cache hit-rate and memory controller requests and average latency.
For these metrics, we can plot various \textbf{statistics} such as the average, maximum and minimum values across all tiles as well as boxplot and standard deviation, to observe the work distribution throughout the execution.
This time series also provides insights into the length and challenges of the tail of the execution, where the maximum and minimum values are far apart.
In addition to plotting statistics about these performance counters, the GUI can also generate a \textbf{heatmap} of the tile grid, where the color represents the percentage of the duration of the frame ($\sim$microseconds) that the counter was activated.
The frames are then played in sequence (by creating a GIF) to visualize the evolution of that metric on the tile grid over time.
For example, \cref{fig:mesh_heatmap} shows that for the routing activity when using a mesh (top), a torus (middle), and a torus with reduction trees (bottom).
The left panels show the routing activity and the right panels show the PU (core) activity.
Overall, \proj's visualization tool helps identify the bottlenecks of different applications and datasets, and how they change with different DUT configurations.

\textbf{\textit{Verbosity:}}
\proj supports four levels of output verbosity: (v=0) only reports aggregated statistical metrics at the end of the execution; (v=1) reports these metrics for each time frame; (v=2) reports the metrics for each tile in the grid, which is required to plot heatmaps; (v=3) also shows the capacity of the input and output queues of each task type for every tile.
Higher verbosity levels increase the log file size and the simulator runtime, proportionally to the configured frame rate.

\begin{figure}
\centering
\resizebox{\columnwidth}{!}{
\begin{tabular}{c c}
\subf{\animategraphics[width=110mm]{3}{figs/HEAT0_2_64_Kron22_Router_Active/heatmap_Router_Active_frame_}{1}{49}}
{}
&
\subf{\animategraphics[width=110mm]{3}{figs/HEAT0_2_64_Kron22_Core_Active/heatmap_Core_Active_frame_}{1}{49}}
{}
\\
\subf{\animategraphics[width=110mm]{3}{figs/HEAT1_2_64_Kron22_Router_Active/heatmap_Router_Active_frame_}{1}{27}}
{}
&
\subf{\animategraphics[width=110mm]{3}{figs/HEAT1_2_64_Kron22_Core_Active/heatmap_Core_Active_frame_}{1}{27}}
{}
\\
\subf{\animategraphics[width=110mm]{3}{figs/HEAT5_2_64_Kron22_Router_Active/heatmap_Router_Active_frame_}{1}{15}}
{}
&
\subf{\animategraphics[width=110mm]{3}{figs/HEAT5_2_64_Kron22_Core_Active/heatmap_Core_Active_frame_}{1}{15}}
{}
\end{tabular}
}
\vspace{-1mm}
\caption{
Animation of the router activity when running BFS on RMAT-22 for three different NoCs: 2D mesh (top), 2D torus (middle), and 2D torus with reduction trees (bottom).
The left panels show the routing activity and the right panels show the PU (core) activity.
No router activity can mean that the router has no messages to route, or that the NoC is clogged and messages are stuck.
The animation is composed of snapshots at a rate of a frame every 40 microseconds (this rate is configurable in \proj).
Since this rate is the same for these plots, the number of frames (50, 28, and 16, from top to bottom) is proportional to the execution time.
The animation can be visualized by opening this PDF with Adobe.
Visualizing the router and core activity simultaneously helps understand the effect of NoC congestion on core utilization.
In addition, plotting the destination-port collisions helps understand the router activity.
We evaluated the version of BFS with barrier synchronization at the end of each epoch to showcase the effect of the network on the tail execution time (3 major epochs can be observed).
A finer time resolution allows observing the evolution of the execution in more detail, but it increases the size of the GIF.
}
\vspace{-3mm}
\label{fig:mesh_heatmap}
\end{figure}

\subsection{Benchmark Suite}\label{sec:apps}

\textbf{\textit{Applications:}}
\proj includes four graph algorithms, two sparse linear algebra, and two HPC kernels~\cite{parboil}.
\textit{Breadth-First Search (BFS)} determines the number of hops from a root vertex to all vertices reachable from it;
\textit{Single-Source Shortest Path (SSSP)} finds the shortest path from the root to each reachable vertex;
\textit{PageRank} ranks websites based on the potential flow of users to each page~\cite{pagerank};
\textit{Weakly Connected Components (WCC)} finds and labels each set of vertices reachable from one to all others in at least one direction (using graph coloring~\cite{connected_components});
\textit{Sparse Matrix-Vector Multiplication (SPMV)} multiplies a square sparse matrix with a dense vector.
\textit{Sparse Matrix-Matrix Multiplication (SPMM)} multiplies a square sparse matrix with a dense matrix and stores the result in a dense matrix.
\textit{3D Fast Fourier Transform (FFT)} computes the Fourier Transform of a 3D tensor.
\textit{Histogram} counts the values that fall within a series of intervals.
The implementation of BFS, SSSP and WCC, supports running with local or global barrier synchronization at the end of each epoch, where the next vertices to explore are stored in the frontier.

As throughput, \proj reports FLOPS (considering the dataset as arrays of FP32) and traversed edges per second (TEPS) as $TEPS=m/time$ where $m$ is the number of edges connected to the vertices in the graph starting from the search key.
When reporting TEPS for SPMV, SPMM, FFT and Histogram, \proj considers the non-zero elements to multiply, and elements to process, respectively.

\textbf{\textit{Datasets:}}
\proj includes ten datasets, including six sizes of the RMAT~\cite{kron} graphs---standard on the Graph500 list~\cite{graph500}---RMAT-16, 21, 22, 25, 26, and 27, which are named after their scale.
For example, RMAT-26 contains $2^{26}$, i.e., 67M vertices (V) and 1.3B edges (E), and has a memory footprint of 12GB.
We also include the real-world graphs from Wikipedia (V=4.2M, E=101M), LiveJournal (V=5.3M, E=79M), Amazon (V=262K, E=1.2M) and Twitter (V=81K, E=2.4M)~\cite{snapnets}.
For SPMV and SPMM, the graphs are seen as a square sparse matrix with V rows and columns and E non-zero elements.
The graphs (as sparse matrices) are stored in the Compressed Sparse Row (CSR) format without any partitioning, i.e, the dataset contains three input arrays, one for the values of the non-zeros, one for the column indices of those non-zeros, and one for the pointers to the beginning of each row in the previous two arrays.
For Histogram, the output array is the count of the column indices of the non-zeros.

\section{Results}\label{sec:results}

This section first presents our validation of \proj by simulating the Cerebras Wafer Scale Engine (WSE)~\cite{cerebras_hotchips} and comparing the results with an existing experimental evaluation~\cite{cerebras_fft}.
Then, \cref{sec:results_speed} shows the speedup of \proj with the number of host threads and its throughput when simulating up to a million tiles.
Finally, \cref{sec:use_case} presents a case study of \proj to explore the optimal memory integration for a given application domain, for several target metrics.

Our \repo includes additional experiments and case studies that explore the impact of
(1) the NoC topology, width and frequency, (2) the reduction-tree size (3) the number of PUs per tile and the PU frequency, (4) integrating DRAM or not, which sets the minimum parallelization required for a given dataset, and (5) the size of the tile input and output queues, among others.

\subsection{Validation}\label{sec:results_validation}

We compare the runtimes reported on the WSE for Fast Fourier Transforms (FFTs)~\cite{cerebras_fft} with the runtimes obtained with our simulator.
We perform simulations for all their reported datapoints, i.e., parallelizing FFT of $n^3$ tensors across $n^2$ processors, for $n$ equal to 32, 64, 128, 256, and 512.
In doing so, we show that \proj has a high level of scalability and the accuracy is not impacted by the size of the DUT.

The WSE mounted on the CS-2~\cite{cerebras_hotchips} is a 46,225mm$^2$ monolithic die with 850,000 cores and 40GB of SRAM on 7 nm technology.
To model the WSE, we configure the DUT as a package with a single chiplet (32-bit 2D mesh NoC) with no integrated DRAM~\cite{cerebras_hotchips}.
The WSE has a circuit-switched NoC with traffic filtering, while our simulator models a packet-switched NoC.
The advantage of the circuit-switched NoC is to avoid message headers and simplify the routers, but it requires more synchronization when setting up a path to send data.
To overcome this difference, we do not consider the destination header in this case, and rather consider the same amount of data sent by the PUs in both cases.
The performance model for the PUs is set based on their reported numbers~\cite{cerebras_fft}.

\textbf{\textit{Runtime:}}
The runtimes of the $n^3$ Fast Fourier Transform (FFT) problems parallelized across $n^2$ tiles reported on the WSE~\cite{cerebras_fft} are $1.2\times$ of the runtimes reported by \proj, consistently, for $n$ ranging from 32 to 512 (i.e., from a thousand to a quarter million tiles).
We believe the reason for this slightly optimistic runtime is that \proj does not model the synchronization overhead of the circuit-switched NoC of the WSE.
We also simulated these datapoints with up to 32 host threads to measure the speedup of the simulator, shown in \cref{fig:sim_time} for FFT and other applications.

In terms of \textbf{\textit{chip area}}, the area reported by the simulator is 8.8\% larger than the area of the WSE.
We attribute this to WSE's NoC, where their routers are likely smaller than the default area of \proj's routers.
Regarding \textbf{\textit{energy}}, the simulator reports an average power of the processing tiles of a bit over 1 KW.
While the WSE study did not report energy for the FFT evaluation, we know that the entire CS-2 system (including cooling) draws up to 20 KW~\cite{cerebras_hotchips}.
Considering that 512x512 is less than a third of the CS-2, and that PU utilization is low ($\sim$30\%) because of the communication bottleneck of FFT, the energy estimation seems reasonable.

\subsection{Parallelization Speedup and Scaling Throughput}\label{sec:results_speed}

\begin{figure}[t]
\centering  
\includegraphics[width=\columnwidth]{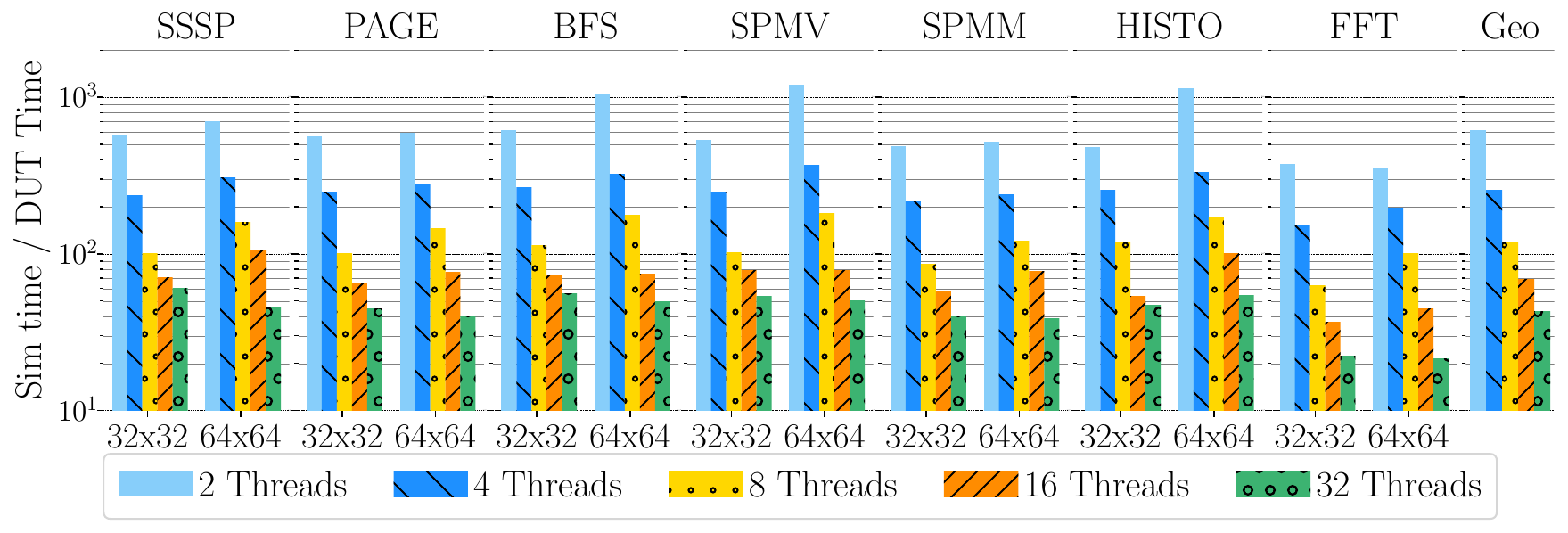}
\caption{
Ratio between the simulator and the DUT runtime, for two DUT sizes (32x32 and 64x64 tiles, monolithic, connected via a 64-bit 2D torus), evaluated with an increasing number of host threads to process the same RMAT-22 dataset.
The DUT time is considered as the aggregated runtime of all tiles.
The simulator runtime decreases close to linearly with the number of host threads.
}
\label{fig:sim_time}
\end{figure}

\cref{fig:sim_time} shows the performance of \proj with the number of host threads, for two DUT sizes. 
This is shown as the ratio between the host runtime for the simulator and the simulated runtime of executing these applications on the DUT runtime.
This ratio decreases from a geomean ratio of 614 to 43 (a 12$\times$ speedup) across the datapoints, when scaling from 2 to 32 host threads (Intel Xeon Gold 6342 at 2.8Ghz).
The speedup in \proj's parallelization is linear until the point where each host thread only processes a couple of tile columns (32-thread 32x32 DUT).
\cref{fig:sim_time} demonstrates the efficiency of \proj and its parallelization.
On a 32-thread host, simulating the execution of applications on the DUT takes only $43\times$ (on geomean, and down to $21\times$ for FFT) longer than the expected runtime of every tile of the DUT.
This ratio is remarkable considering that \proj performs a functional simulation of data-dependent workloads, where all NoC routers are evaluated every cycle at the flit level.

\begin{figure}[t]
\centering  
\includegraphics[width=\columnwidth]{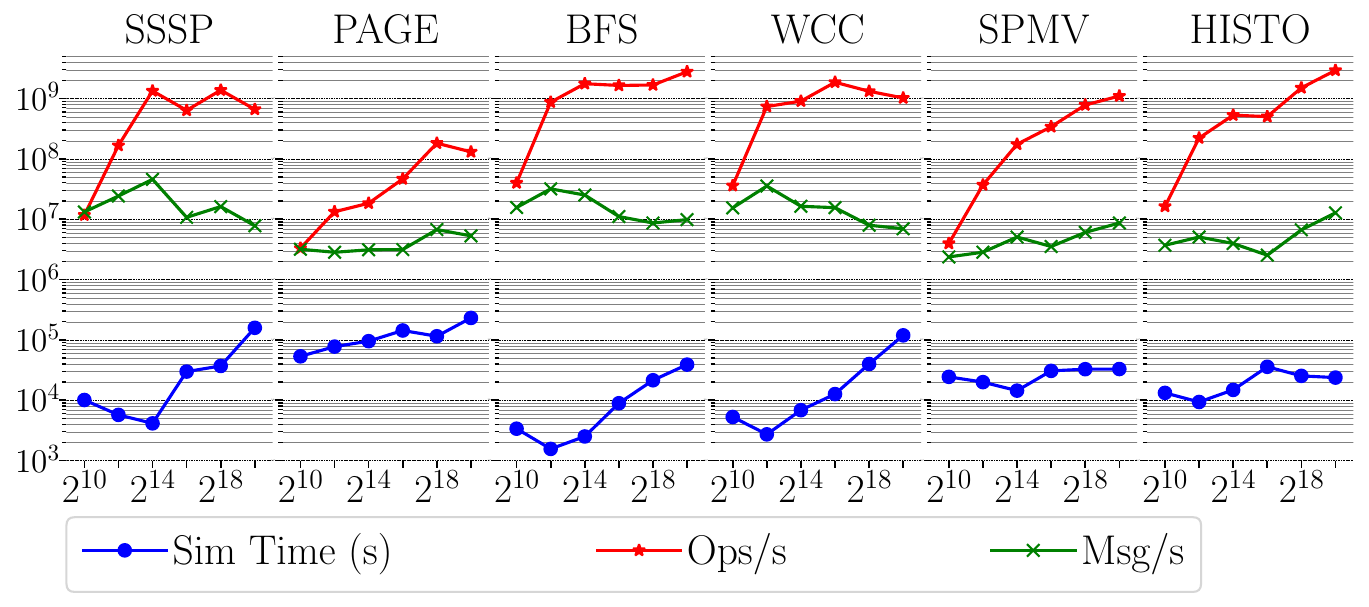}
\caption{
Simulation time (in host seconds) and throughput in DUT operations and NoC message fits routed per second (y-axis), for scaling DUT sizes from a thousand to a million tiles (x-axis) when processing the RMAT-26 dataset.
(This evaluation models 32x32 tiles per chiplet, connected via a 64-bit hierarchical 2D torus.)
The $2^{10}$ and $2^{12}$ datapoints are evaluated with 16 and 32 host threads, respectively, of a single-socket Intel Xeon Gold 6342 at 2.8Ghz, the datapoints from $2^{14}$ to $2^{18}$ with 64 and $2^{20}$ with 128 threads, respectively, of a 4-socket Intel Xeon Gold 6230 at 2.1GHz.
}
\label{fig:scaling}
\end{figure}

\cref{fig:scaling} shows the simulation time and throughput of \proj (in host seconds) when using increasingly large DUT configurations (from a thousand to a million tiles) to execute the RMAT-26 dataset.
The first three datapoints see a decreased or nearly constant simulation time because we are increasing the parallelization of \proj from 16 to 64 host threads.
The NoC simulation throughput ranges from a few million message flits routed per second (PageRank) to over 40 million (SSSP).
This NoC throughput does not count the routing attempts that fail due to contention (two input ports compete for the same output) or backpressure (the buffer of the output port is full).
The throughput of operations---natively executed on the host---reaches up to a few billion Ops/s.

The absolute runtimes in \cref{fig:scaling} (blue) showcase that even million-tile DUT evaluations on a billion-element dataset---unattainable before \proj---are simulated in under half a day for BFS, SPMV and Histo, and in up to two days for SSSP and PageRank, on a single host server.

FFT is not shown in \cref{fig:scaling} since for this benchmark we scale the problem size with the system size (i.e., the FFT of a $n^3$ tensor is parallelized across $n^2$ processors) but we mention some runtimes here.
FFT is simulated in under 100 seconds for up to $n=128$ (i.e, $2^{14}$ tiles) and it takes around a day to simulate $n=1024$ (a million tiles on a billion-element tensor), where the all-to-all communication of FFT dominates the runtime.

\subsection{Case Study: Memory Integration}\label{sec:use_case}

\begin{figure}[t]
\centering  
\includegraphics[width=\columnwidth]{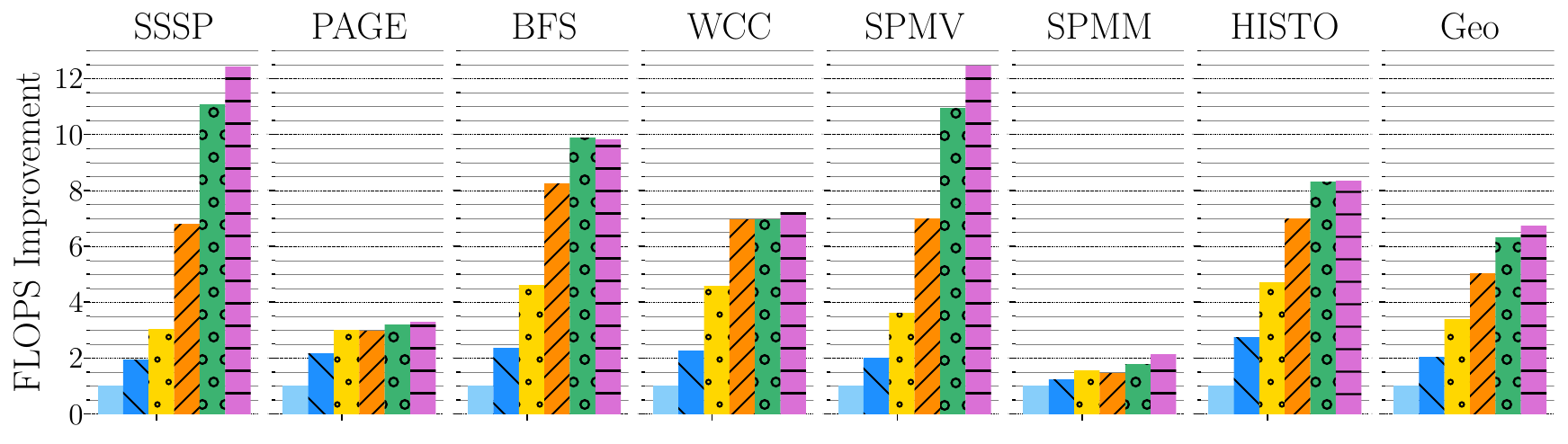}
\includegraphics[width=\columnwidth]{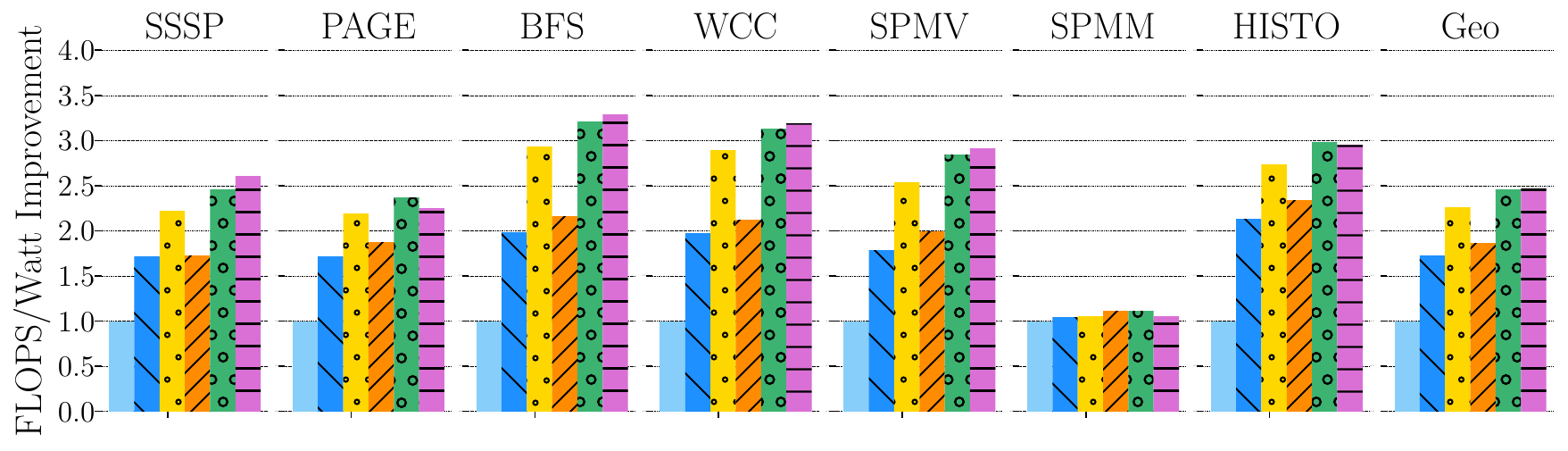}
\includegraphics[width=\columnwidth]{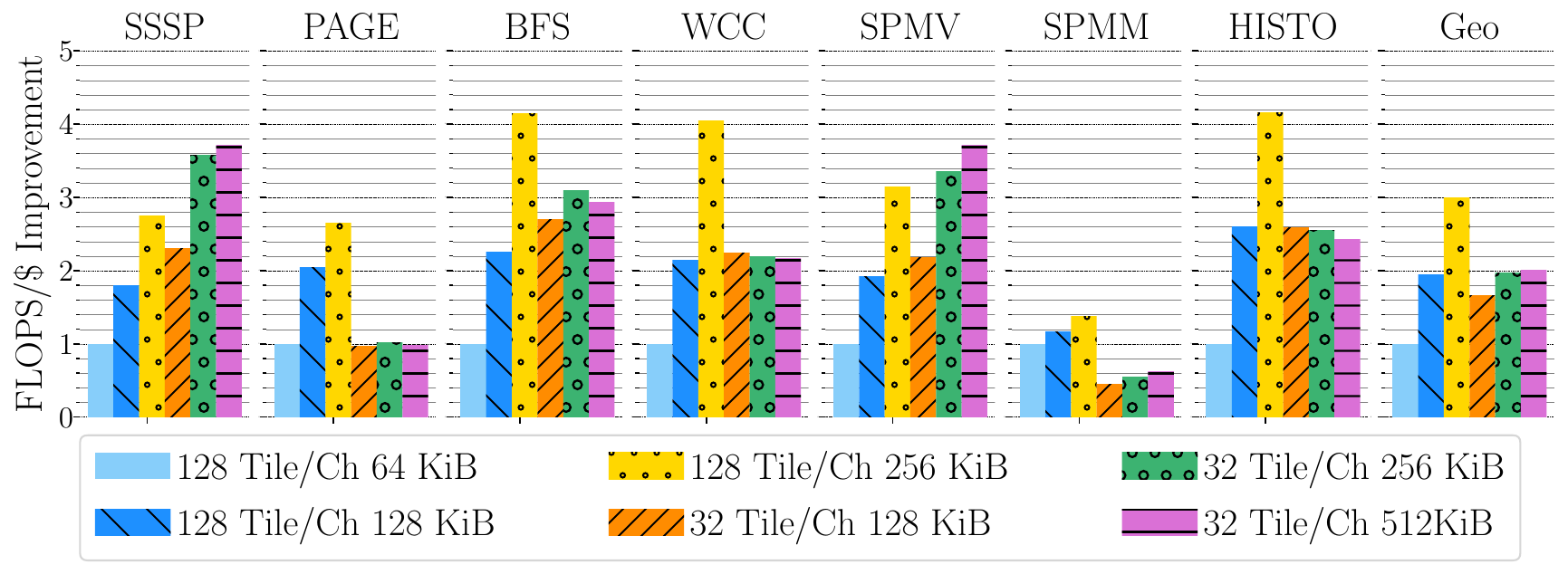}
\vspace{-2mm}
\caption{Performance, energy efficiency, and performance per dollar improvements of the DUT using different SRAM sizes and number of tiles per HBM channel, over a baseline of 64 KiB SRAM and 128 tiles per HBM channel (Tile/Ch).
In this study, a chiplet is always attached to a single 8-channel HBM device, and thus, the number of tiles per chiplet (16x16 or 32x32) determines the ratio of tiles per HBM channel.
The RMAT-25 dataset is studied on a DUT with 1024 tiles; the dataset footprint per tile ranges from 4 MiB (Histogram) to 8 MiB per tile (SPMV).
}
\label{fig:use_case_mem}
\end{figure}

As a case study, we studied the performance, energy efficiency, and performance per dollar improvements of different SRAM sizes and numbers of tiles per chiplet (same number of tiles in total) and show the results in \cref{fig:use_case_mem}. Since all applications but SPMM have a very low \textbf{arithmetic intensity} with respect to the data movement, they require a large amount of memory bandwidth.
The top plot of (\cref{fig:use_case_mem} shows a strong performance increase with SRAM size, $3.5\times$ on geomean when increasing from 64 KiB to 256 KiB, with the same chiplet configuration of 32x32 tiles, and an additional $2\times$ increase when using 16x16-tile chiplets (32 Tile/Ch), for the same SRAM size of 256 KiB.
The \textbf{hit-rate} of the data cache (not displayed) increases from 83\% to 95\% on geomean with the SRAM size increase, but since the effective bandwidth of a tile is $SRAM\_bandwidth \times hit\_rate + DRAM\_bandwidth \times (1 - hit\_rate)$, the hit-rate has a larger impact on the bandwidth when more tiles share the same DRAM channel (128 Tile/Ch).

The \textbf{throughput per dollar} is lower with 16x16-tile chiplets for all applications but SSSP and SPMV, due to the extra cost of having four times more HBM devices.
The throughput per dollar estimates for SPMM---with over an order of magnitude more arithmetic intensity than the rest of the applications for the dataset evaluated (RMAT-25)---showcases that when targeting performance per dollar, the optimal memory integration depends on the application domain.

\section{Conclusions}\label{sec:conclusions}

\proj is a simulation framework aimed to explore the design space of scale-out architectures for communication-intensive applications like graph analytics and sparse linear algebra.
It supports simulating a class of tiled manycore architectures with different hierarchical organizations, network topologies, and memory integrations.
It also supports evaluating various parallelization strategies (do-all and task-based) and communication primitives (e.g., message-passing and reduction trees).
As demonstrated via a case study, \proj is well suited to explore the optimal balance between resources dedicated to compute, memory and network, for different application domains and target metrics (e.g., performance, power and cost).

The distinguishing factor of \proj is its scalability in the challenging domain of data-dependent and communication-intensive applications, which requires functional simulation (since the execution depends on the data) as well as modeling the communication cycle by cycle.
To demonstrate that, we simulated applications with billion-element datasets parallelized with up to a million processing units and presented the simulation times and throughput metrics.
We achieve this scalability by optimizing the design and parallelization of the simulator for the target class of tiled manycore architectures.

The \proj framework is open-source and available at our \repository, including the simulator, applications, datasets, and visualization tools.
It also contains the scripts to reproduce the experiments presented in this paper and other case studies, as a tutorial to allow researchers to explore the design space of scale-out manycore architectures.
Future work on the simulator itself includes multi-node MPI parallelization and support for low-diameter cluster interconnects, to enable simulating even larger systems.